\documentstyle[12pt,aasms4,psfig,epsfig]{article}

\newcommand{\COd}     {\mbox{$^{12}$CO$\;$}}
\newcommand{\COt}     {\mbox{$^{13}$CO$\;$}}  

\received{** 1999} \accepted{** 1999} \journalid{337}{15 January 1989}
\articleid{11}{14} \slugcomment{}

\begin{document}

\def\th{$^{13}$}
\def\ei{$^{18}$}
\def\tw{$^{12}$}
\def\Lcs{{\hbox {$L_{\rm CS}$}}}
\def\lunits{K\thinspace \kms\thinspace pc$^2$}

\def\,{\thinspace}
\def\etal{et al.}


\def\kms{km\thinspace s$^{-1}$}
\def\Lsun{L$_\odot$}
\def\Msun{M$_\odot$}
\def\ms{m\thinspace s$^{-1}$}
\def\percc{cm$^{-3}$}

\font\sc=cmr10

\def\CBR{{\rm\sc CBR}}
\def\FWHM{{\rm\sc FWHM}}
\def\HI{{\hbox {H\,{\sc I}}}}
\def\HII{{\hbox {H\,{\sc II}}}}


\def\Ha{H$\alpha$}                      
\def\Htwo{{\hbox {H$_2$}}$\;$}
\def\nHtwo{n(\Htwo)}
\def\He#1{$^#1$He}                      
\def\water{H$_2$O$\;$}
\def\flecha{\rightarrow}
\def\COJ#1#2{{\hbox {CO($J\!\!=\!#1\!\rightarrow\!#2$)}}}
\def\CO#1#2{{\hbox {CO($#1\!\rightarrow\!#2$)}}}
\def\CeiO#1#2{{\hbox {C$^{18}$O($#1\!\rightarrow\!#2$)}}}
\def\CSJ#1#2{{\hbox {CS($J\!\!=\!#1\!\rightarrow\!#2$)}}}
\def\CS#1#2{{\hbox {CS($#1\!\rightarrow\!#2$)}}}
\def\HCNJ#1#2{{\hbox {HCN($J\!\!=\!#1\!\rightarrow\!#2$)}}}
\def\HCN#1#2{{\hbox {HCN($#1\!\rightarrow\!#2$)}}}
\def\HNC#1#2{{\hbox {HNC($#1\!\rightarrow\!#2$)}}}
\def\HNCO#1#2{{\hbox {HNCO($#1\!\rightarrow\!#2$)}}}
\def\HCOpJ#1#2{{\hbox {HCO$^+$($J\!\!=\!#1\!\rightarrow\!#2$)}}}
\def\HCOp#1#2{{\hbox {HCO$^+$($#1\!\rightarrow\!#2$)}}}
\def\HCOpp{{\hbox {HCO$^+$}}}
\def\J#1#2{{\hbox {$J\!\!=\!#1\rightarrow\!#2$}}}
\def\noJ#1#2{{\hbox {$#1\!\rightarrow\!#2$}}}


\def\Lco{{\hbox {$L_{\rm CO}$}}}
\def\Lhcn{{\hbox {$L_{\rm HCN}$}}}
\def\Lfir{{\hbox {$L_{\rm FIR}$}}}
\def\Ico{{\hbox {$I_{\rm CO}$}}}
\def\Sco{{\hbox {$S_{\rm CO}$}}}
\def\Ihcn{{\hbox {$I_{\rm HCN}$}}}


\def\Tastar{{\hbox {$T^*_a$}}}
\def\Tmb{{\hbox {$T_{\rm mb}$}}}
\def\Tb{{\hbox {$T_{\rm b}$}}}

\title{O-bearing Molecules in Carbon-rich Proto-Planetary Objects\footnotemark[1]}

\author {F. Herpin\footnotemark[2], J. Cernicharo}


\affil {Depto F\'{\i}sica Molecular, I.E.M., C.S.I.C, 
Serrano 121, E-28006 Madrid, Spain}



\footnotetext[1]
{ Based on observations with ISO,
an ESA project with instruments funded by ESA Member States
(especially the PI countries: France, Germany, the Netherlands
and the United Kingdom) and with participation of ISAS and NASA.}
\footnotetext[2]{herpin@isis.iem.csic.es}

\begin{abstract}

We present ISO LWS observations of the proto-planetary nebula CRL 618, a star
evolving very fast to the planetary nebula stage. In addition to
the lines of \COd, \COt, HCN and HNC, we report on the detection
of \water and OH emission together with the fine structure lines of
[OI] at 63 and 145 $\mu$m. The abundance of the latter three 
species relative to \COd are 4 $10^{-2}$, 8 $10^{-4}$ 
and 4.5 (approximate value) in the regions where they are
produced. We suggest that O-bearing species other than CO are produced
in the innermost region of the circumstellar envelope. The UV photons 
from the central star photodissociate most of the molecular species
produced in the AGB phase and allow a chemistry dominated
by standard ion-neutral reactions. Not only allow these reactions the
formation of O-bearing species, but they also modify the abundances of C-rich
molecules like HCN and HNC for which we found an abundance ratio of 
$\simeq$ 1, much lower than in AGB stars.
The molecular abundances in the different regions of the circumstellar
envelope have been derived from radiative transfer models and our knowledge
of its physical structure.

\end{abstract}


\section{Introduction}

CRL 618 is one of the few 
clear examples of an AGB
star in the transition phase to the Planetary Nebula stage. It has 
a compact HII region
created by a hot central star (Westbrook 1976, Kwok \& Feldman 1981). 
This object is observed as a bipolar nebula at 
optical, radio and infrared wavelengths (Carsenty \& 
Solf ~1982, Bujarrabal \etal ~1988, and Hora \etal ~1996). 
The expansion velocity of the envelope
is around 20 \kms, but CO observations  
show the presence of a high-velocity outflow with velocities up 
to 300 \kms (Cernicharo \etal ~1989). High-velocity emission in the 
\Htwo $v=1-0$ S(1) line is also detected (Burton \& Geballe 1986).
The high velocity wind and the UV photons from the star 
perturb the circumstellar envelope (CSE) 
producing shocks and photodissociation regions (PDRs) 
which modify the physical and chemical conditions of the gas
(Cernicharo \etal ~1989, and Neri \etal ~1992). 

We present in this Letter an ISO (Kessler \etal ~1996) 
Long Wavelength Spectrometer 
(LWS, Clegg \etal ~1996, Swinyard \etal ~1996) 
spectrum between 43 and
197 $\mu$m. In addition to the high-J lines of CO, HCN and HNC, we have 
discovered, for the first time in a C-rich star, several lines of water vapor, 
OH and the fine structure lines of [OI]. We discuss the origin of these 
O-bearing species and we present a model that accounts for 
the observed emission. 

\section{Observations and Results}

The LWS data were taken on revolution 688. 
The data have been processed following pipeline number 7. We have used 
ISAP\footnotemark[3]    to remove glitches and fringes. The rotational lines 
(see Fig. 1 and 2) of \COd (J=14-13 to J=41-40),
of \COt (J=14-13 to J=19-18), HCN, HNC, [OI] (at 63.170 and 
145.526 $\mu$m) and C$^{+}$ ($^{2}p_{3/2}- ^{2}p_{1/2}$ transition at 
157.74 $\mu$m) are observed. 
Several lines of OH and \water are also detected (see Fig. 1 and 2). 

\footnotetext[3]{The ISO Spectral Analysis Package (ISAP) is a joint development 
by the LWS and SWS Instrument Teams and Data Centers. 
Contributing institutes are CESR, IAS, IPAC, MPE, RAL and SRON.}  

The far-infrared spectrum is essentially
dominated by the CO lines. All the other species have much lower intensities
contrary to what is found in the AGB star IRC+10216 where the HCN lines
(from the ground and vibrationally excited states) are as strong as CO.
The main difference is due to the physical structure of the CSE of 
CRL 618  which has a central hole in molecular species filled by 
a bright HII region.
The LWS data also show several weak lines that remain unidentified.
We have searched for CH$^+$, NH, CH, and other light species. But unlike the
case of NGC7027 (Cernicharo et al. ~1997) where the PDR has a large spatial
size, the PDR of CRL 618 is still close to the central object and has a
larger dilution in the LWS beam, which makes the fluxes of these expected 
species very weak. 

The full spectrum of CRL 618 between 2 and 200 $\mu$m has been discussed
by Cernicharo \etal ~(1999a et 1999b). In these data, in addition
to the species detected in this Letter, they have also reported
the presence of several long polyacetylenic chains and the bending modes of
HC$_3$N and HC$_5$N. However, in the LWS part of the spectrum no pure
rotational lines of HC$_3$N have been found. Nevertheless, the $\nu_7$ band
of HC$_3$N at 44 $\mu$m is observed and these data will be published
elsewhere. All the observed absorption bands in the mid-infrared arise
from gas at $\simeq$ 250 K and densities larger than $10^{7}$ cm$^{-3}$ 
(Cernicharo \etal ~1999b).

\section{Discussion}

Previous works on CRL 618 underlined the presence of a central 
torus associated with a 
Photo-Dissociated Region, an extended AGB remnant envelope 
and high velocity wind regions (HVW) as bipolar lobes 
(Weintraub \etal ~1998, Burton \& Geballe 1986, Neri \etal ~1992, 
Carsenty \& Solf 1982), whose 
physical parameters, like the velocity of the gas or the angular size, were investigated 
(Martin-Pintado \etal ~1995, Meixner \etal ~1998, Hajian \etal ~1996). 
In order to reproduce the emission, we have 
to introduce several spatial components, in terms of kinematics, temperature 
and density based on the 
knowledge we have of this object. In Fig. 3 we show the adopted geometry: 
(i) a central torus with the PDR; (ii) the extended AGB remnant envelope; 
(iii) the lobes (bipolar jets, HVW). 
We tried to reproduce the LWS observations by simulating 
a spectrum (\COd+\COt+HCN+\water+OH+[OI]). 
We first tried to model the \COd intensities. 
Then we applied the same parameters to the other molecules 
allowing to vary only these column densities (see parameters in Table 1). 
We used a simple LVG-model and obtained satisfactory fits for several 
molecules (see Fig. 1). 

The torus is assumed to be composed of a
large neutral shell of 1.1''-1.5'' size,
temperatures from 250 to 800 K, density of  $5 \;10^{7}$ cm$^{-3}$,
and an expansion velocity of 20 \kms. All the detected
molecules are present in the 250 K area where most of the CO long wavelength
emission comes from. In this region the \COd column density ({\em N})
is $10^{19}$ cm$^{-2}$ and
the \COt, \water, HCN, HNC and OH abundances relative
to \COd are $1/20$, $1/25$, $1/1000$, $1/1000$  and $1/1250$ respectively.
We adopt an ortho-to-para ratio for \water around 3.
A warmer region in the torus is needed to reproduce the \COd and \COt emissions
at shorter wavelengths  (size $\simeq 0.6'', $N$=1 \;10^{19}$ cm$^{-2}$,
T=1000 K) and the atomic oxygen emission.
As we reach this inner edge of the PDR, only the atomic oxygen seems to
be present, whereas all molecular species are
photodissociated: $N$ ([OI])$=$4.5 $10^{19}$ cm$^{-2}$, thus an abundance
of 4.5 relative to CO. But this value must be considered as very approximate
due to the uncertainties in the size of this region. For the inner radius of
the torus, which should be in contact with the HII region, we calculate at
a distance of $0.3''$ ($10^{16}$ cm) a UV flux of 20 000 G$_{0}$ (in units of 
the local IS radiation flux); this UV illumination is around 200 in the 
lobes (Latter \etal 1992).

The contribution from the bipolar lobes corresponds to the high velocity gas
(200 \kms, Cernicharo \etal ~1989).
Previous observations of CO line wings (Cernicharo \etal ~1989)
and \Htwo lines (Herpin \etal ~in preparation) 
indicate high velocity and
dense shock-heated molecular gas in CRL 618. The contribution of such a region
of dense warm shocked gas could dominate the CO high rotational transitions
(e.g., Hollenbach \& McKee 1989). The 
high temperature component at high velocity is necessary to produce 
part of the emission at short wavelengths, since a high temperature 
component arising from the torus produces too weak intensities. 
We assume that the bipolar
lobes in CO have a size between 1.5'' and 1.7'' 
with {\em N} ranging from $2 \;10^{16}$ to
5 $10^{18}$ cm$^{-2}$, temperatures of 200 K and 1000 K, 
and density around $10^{7}$  cm$^{-3}$. A contribution from this HVW 
region for \water, OH and [OI] line emission is likely because the molecules 
in the AGB remnant must be dissociated in the shocks, and in the molecular 
recombination some O-bearing species could be found like in the inner part 
of the torus. However, due to the lack of velocity resolution, we can not 
discriminate both components.

The contribution of the extended AGB remnant envelope is as a colder zone 
(100 K) for \COd and \COt.
The density is around 5 $10^{5}$  cm$^{-3}$ and can be 
compared to what was found by
Yamamura \etal ~(1994) with a mass loss rate of $10^{-4}$ M$_{\sun}/$yr for
CRL 618, and Neri \etal ~(1992) ($10^{5}-10^{6}$ cm$^{-3}$ on the outer edge
of the lobes). 

For the HVW region and the AGB remnant envelope,
we adopt a temperature profile derived from the
AGB phase case (Langer \& Watson 1984, Herpin 1998), going from 1000 K on
the origin of the conical structure (corresponding to the temperature
at $6-7 \;10^{14}$ cm for a Mira at 3000 K) to 100 K on the
exterior, equal to the AGB temperature for dust in the outer part of the
envelope (distance $> 10^{16}$ cm) and equal to the value derived by
Yamamura \etal ~(1994). 

The model fits well all the CO lines, with small discrepancies at short
wavelengths mainly due to the presence of different molecular contributions,
to a lower signal-to-noise ratio, and to the fact that
some CO lines are mixed with \water and OH lines.
Nevertheless,
all the observed line intensities from J=14-13 up to J=41-40 are reproduced.
Our model also fits the observed emissions J=1-0, 2-1
(Cernicharo \etal ~1989),  3-2 (Gammie \etal ~1989) and 6-5 (unpublished data).
A good fit is also obtained for the atomic oxygen lines at 63 $\mu$m
and 145 $\mu$m.

Although the \COt emission is weak, the full coverage of the LWS
spectrum allows us to derive a [\COd/\COt] ratio of 20, lower than the common 
interstellar value of 40-45, but equal to the detected ratio (20) 
of Kahane \etal ~(1992) in this object.

Emission from the HCN wings (Neri \etal ~1992) is produced by the strong impact
of a primary outflow on a few localized dense clumps
in the inner envelope.
According to these authors,
70\% of the HCN line emission comes from a very compact, spherically symmetric
region, and is mainly collisionally excited and strongly affected by
self-absorption. They found for the flow (our HVW region) $n(H_{2}) =
\rm few \;10^{5}-10^{6} \;$cm$^{-3}$, $N$(CO)$=2\;10^{18}$ cm$^{-2}$ and
$N(HCN)= \rm few \;10^{17} \;$cm$^{-2}$, which leads to [HCN/CO] around $10^{-1}$.
Our high-velocity HCN component (see Table 1) comes
from the lobes (HVW)
($N= 2\;10^{15}$ cm$^{-2}$, T$=1000$ K).
We find here a ratio [HCN/CO] of $10^{-1}$ equal to the
ratio obtained by Neri \etal ~(1992). When one considers an abundance of CO of a few
$10^{-4}$, this leads to a HCN abundance of a few $10^{-5}$ for our data
($[$HCN/HNC$] \simeq$ 10).
This is the expected abundance ratio in a C-rich star
in the AGB phase (Cernicharo et al. ~1996), and proves
that this HCN material has not been processed by the UV photons from the
central hot star.
Furthermore, Deguchi \etal ~(1986) found in the disk
surrounding the source an abundance of HCN of
order of 2 $10^{-8}$ per \Htwo molecule,
much less than Neri's value. For this region (see Table 1) we have found a ratio
$[$HCN/CO$]$ of $10^{-3}$ which leads to an abundance of a few $10^{-7}$.
We can conclude that in this region HCN molecules are already
efficiently dissociated. The HCN and HNC abundances are similar, which
implies a high degree of reprocessing of the molecular gas
($\chi(\rm HCN)>>\chi(\rm HNC)$ in standard C-rich AGB stars).

Below 110 $\mu$m our models predict that the \water and OH lines will be 
in emission. However, the data seem to indicate that some of these lines 
could be in absorption or are too weak. Taking into account 
the reduced sensitivity at these wavelengths and the complex structure of 
the region, a more detailed model including radiative transfer 
effects for \water and 
OH, similar to those discussed by Gonzalez-Alfonso \& Cernicharo (1999),
will need a better knowledge of the physical and geometrical structure of 
the innermost region of the envelope.

The detection of the fine structure lines of [OI] at 63 and 145 $\mu$m 
gives important insight into the chemical evolution of CRL 618. 
This indicates that the reservoir of CO molecules formed in the AGB phase
of CRL 618 has started to be reprocessed through UV photons and shocks.
In CRL 618 Cernicharo \etal ~(1989) also found H$_2$CO and Bujarrabal 
\etal ~(1988) found HCO$^+$. These molecules are not formed efficiently 
in the AGB phase of a C-rich star. How are they formed in CRL 618 ?
As soon as some atomic oxygen is released into the gas phase from
the photodissociation of CO, standard ion-neutral reactions can form quickly
H$_3$O$^+$ and HCO$^+$. From the former, and through electronic recombination,
\water and OH are efficiently produced. We expect these molecules to
arise from the innermost region of the envelope where the UV flux is 
largest. In our models, the column densities in that region are 10$^{19}$,
5 10$^{17}$ and 8 10$^{15}$ cm$^{-2}$ for \COd, \water
and OH respectively.
That means that the abundances of \water and OH relative to CO are
$4\; 10^{-2}$ and $8\; 10^{-4}$ respectively. These abundances are much
lower that those found in the surrounding media of galactic HII regions
(Cernicharo \etal ~1999a) and suggest that only a small fraction of the UV photons
from the central star are reaching the region where \water and OH are
formed. In fact, we could expect a dusty region in the torus around the
central HII region that protects partially the molecules formed in the AGB
phase. Obviously, in the PDR itself all molecules could be destroyed and
most of the [OI] dominant coolant in high density PDRs emission 
is probably coming from that region.

The model derived from the LWS spectra leads to two global regions of molecular 
emission. One in the torus with several molecular species 
(\COd, \COt, HCN, HNC, \water and OH) at moderate 
temperature. Another in the lobes dominated by CO and HCN molecules.
The detected atomic oxygen 
line is strong and comes from the innermost warm region, the inner edge of the PDR, 
what proves that CRL 618 is already quite well on its way to the PN stage and that 
these objects, even in the C-rich case, can produce efficiently O-bearing molecular 
species.

{\it Acknowledgements}
We thank spanish DGES and CICYT for funding support
for this research under grants PB96-0883 and ESP98-1351E.

\clearpage 
\begin{table*} [h]
  \caption{ \label{table1}Table of the column densities (in cm$^{-2}$) for all
molecules present in each region of the object as depicted in Figure 3 (see 
text for the adopted parameters).}
  {\small{\begin{tabular}{lccccccc} \hline \hline
 {\bf Molecules} & {\bf \COd} & {\bf \COt} & {\bf HCN} & {\bf HNC} & {\bf o-\water}
& {\bf OH} & {\bf O[I]} \\ \hline
{\bf TORUS} (including PDR) & & & & & & & \\
T=250 K, $\theta= 1.5''$ & $10^{19} $ & 5 $10^{17}$ & $10^{16}$
& $10^{16}$ & 3 $10^{17}$ & 8 $10^{15}$ & \\
$V=20$ \kms, $n(H_{2})=$5 $10^{7}$ cm$^{-3}$ & & & & & & & \\
T=800 K, $\theta= 1.1''$ & 6 $10^{17} $ & 3 $10^{16}$ & & & & & \\
$V=20$ \kms, $n(H_{2})=$5 $10^{7}$ cm$^{-3}$ & & & & & & & \\
T=1000 K, $\theta= 0.6''$ & 1 $10^{19} $ & 5 $10^{17}$ & & & & 
& 4.5 $10^{19}$  \\
$V=20$ \kms, $n(H_{2})=$7 $10^{7}$ cm$^{-3}$ & & & & & & & \\  \hline
{\bf HVW} & & & & & & & \\
T=200 K, $\theta= 1.7''$ & 5 $10^{18} $ & 2.5 $10^{17}$ & 3 $10^{17}$
& 2 $10^{17}$ & & & \\
$V=50$ \kms, $n(H_{2})= 10^{7}$ cm$^{-3}$ & & & & & & & \\
T=1000 K, $\theta= 1.5''$ & 2 $10^{16} $ & 1 $10^{15}$ &  2 $10^{15}$
& $10^{14}$ & & & \\
$V=200$ \kms, $n(H_{2})=10^{7}$ cm$^{-3}$ & & & & & & & \\  \hline
{\bf AGB remnant} & & & & & & & \\
T=100 K, $\theta= 5''$ & 7 $10^{18} $ & 3.5 $10^{17}$ & 7 $10^{17}$
& 7 $10^{15}$ & & & \\
$V=20$ \kms, $n(H_{2})=5$ $10^{5}$ cm$^{-3}$ & & & & & & & \\
\hline 
 \end{tabular}}}
\end{table*}

\clearpage
\begin{center}
Figure Captions
\end{center}         

\figcaption[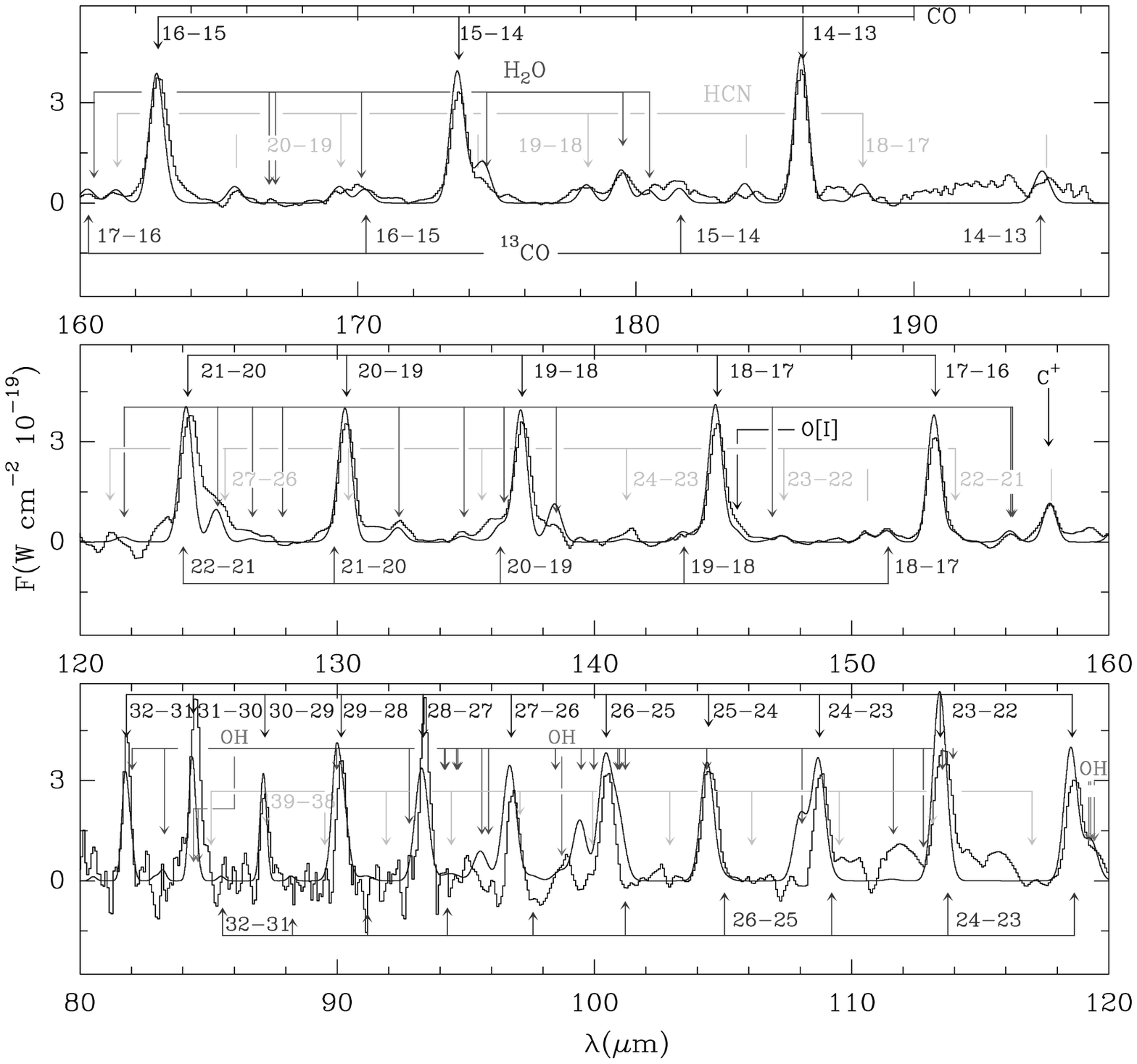]
{Continuum subtracted spectra of CRL 618 (LW detectors). The result
of our model is shown by the continuous line. 
The lines of CO, \COt, HCN, \water and OH are indicated by arrows while
those of HNC are indicated by vertical lines (from J=22-21 at 150.627 $\mu$m 
to J=17-16 at 194.759 $\mu$m). The C$^{+}$ transition is 
not included in our models; the plot indicates a gaussian fit to this feature.}

\figcaption[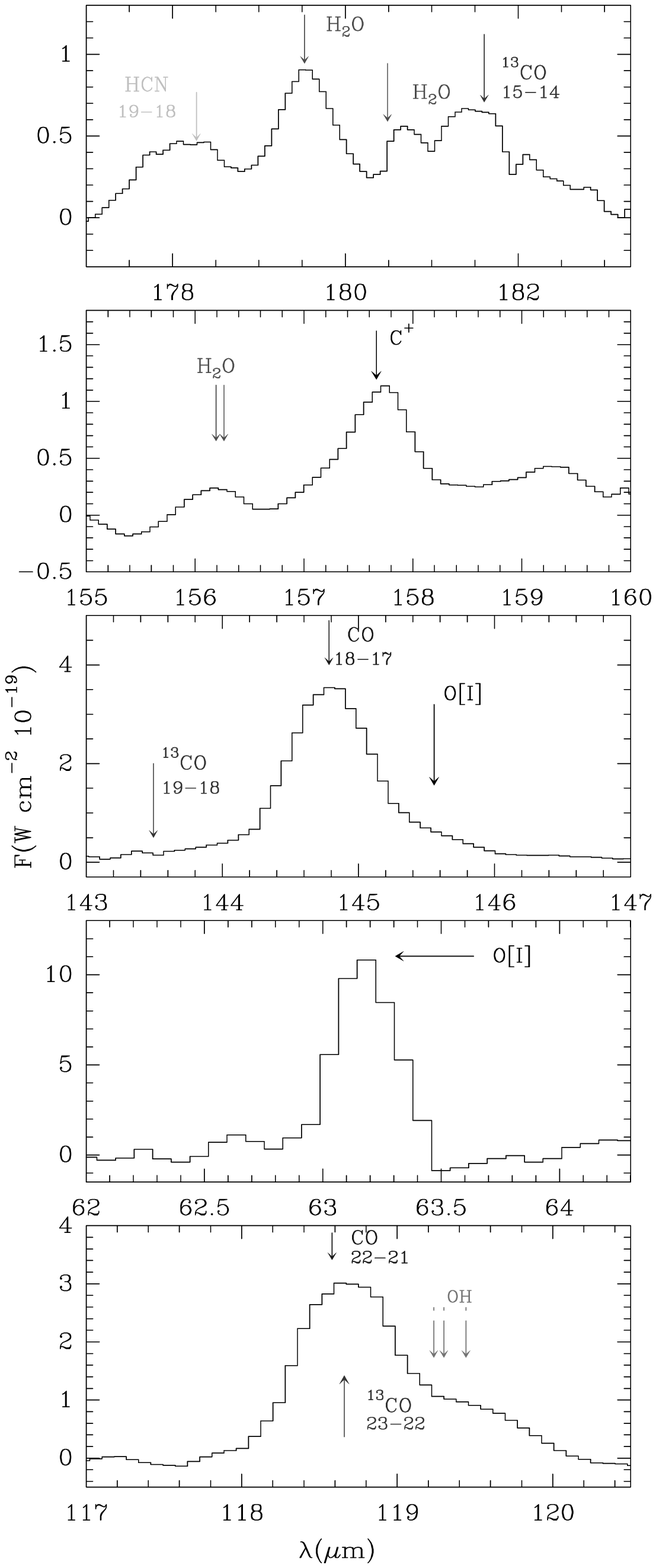]
{Important detected lines in CRL 618 (continuum subtracted spectra).}

\figcaption[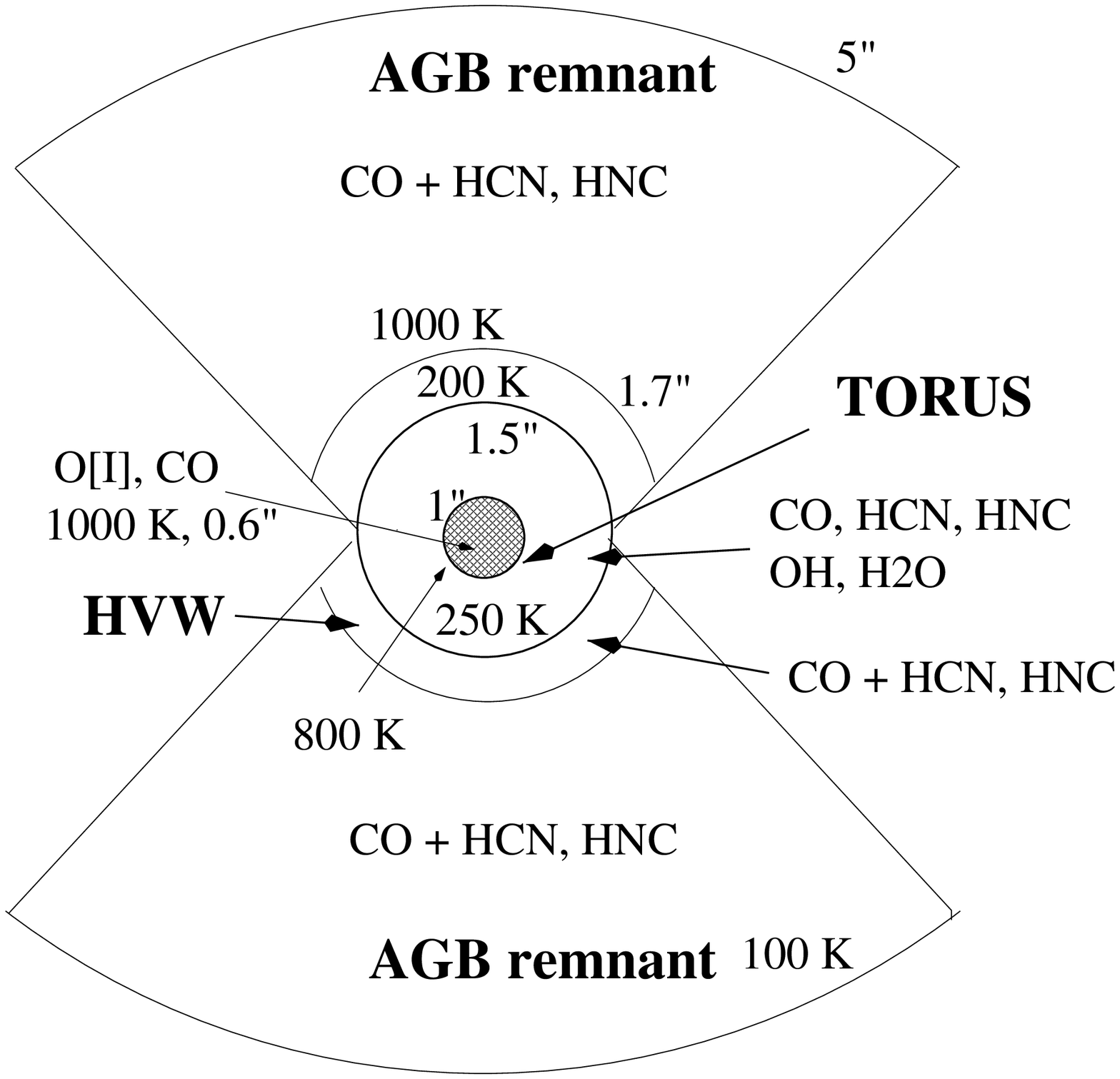]
{Simple view of the molecular and emission structure as seen 
by the model. HVW is for the {\em High Velocity Wind} region.}

\clearpage

\begin{figure}[h]
\plotone{fig1.eps}
\end{figure}

\clearpage                                        

\begin{figure}[h]
\epsfysize=22cm
\epsfxsize=9cm
\plotone{fig2.eps}
\end{figure}

\clearpage

\begin{figure}[h]
\plotone{fig3.eps}
\end{figure}

\end{document}